\documentclass[letter,usenatbib]{mnras}

\usepackage{newtxtext,newtxmath}

\usepackage[T1]{fontenc}
\usepackage{ae,aecompl}

\usepackage{graphicx}	
\usepackage{amsmath}	
\usepackage{amssymb}	

\title[Fe enrichment in ellipticals, groups, and clusters]{Mass-invariance of the iron enrichment in the hot haloes of massive ellipticals, groups, and clusters of galaxies}

\author[F. Mernier et al.]{
F. Mernier,$^{1,2,3}$\thanks{E-mail: mernier@caesar.elte.hu}
J. de Plaa,$^{3}$
N. Werner,$^{1,4,5}$
J. S. Kaastra,$^{3,6}$
A. J. J. Raassen,$^{3}$
L. Gu,$^{7}$
\newauthor
J. Mao,$^{3,6}$ 
I. Urdampilleta,$^{3,6}$
N. Truong,$^{1}$
and A. Simionescu$^{8}$
\\
$^{1}$MTA-E\"otv\"os University Lend\"ulet Hot Universe Research Group, P\'azm\'any P\'eter s\'et\'any 1/A, Budapest, 1117, Hungary\\
$^{2}$Institute of Physics, E\"otv\"os University, P\'azm\'any P\'eter s\'et\'any 1/A, Budapest, 1117, Hungary\\
$^{3}$SRON Netherlands Institute for Space Research, Sorbonnelaan 2, 3584 CA Utrecht, The Netherlands\\
$^{4}$Department of Theoretical Physics and Astrophysics, Faculty of Science, Masaryk University, Kotl\'a\v{r}sk\'a 2, Brno, 611 37, Czech Republic \\
$^{5}$School of Science, Hiroshima University, 1-3-1 Kagamiyama, Higashi-Hiroshima 739-8526, Japan \\
$^{6}$Leiden Observatory, Leiden University, P.O. Box 9513, 2300 RA Leiden, The Netherlands\\
$^{7}$RIKEN Nishina Center, 2-1 Hirosawa, Wako, Saitama 351-0198, Japan \\
$^{8}$Institute of Space and Astronautical Science (ISAS), JAXA, 3-1-1 Yoshinodai, Chuo-ku, Sagamihara, Kanagawa 252-5210, Japan
}

\date{Accepted 2018 May 02. Received 2018 May 02; in original form 2018 March 12}

\pubyear{2018}

\begin{document}
\label{firstpage}
\pagerange{\pageref{firstpage}--\pageref{lastpage}}
\maketitle

\begin{abstract}
X-ray measurements find systematically lower Fe abundances in the X-ray emitting haloes pervading groups ($kT\lesssim1.7$ keV) than in clusters of galaxies. These results have been difficult to reconcile with theoretical predictions. However, models using incomplete atomic data or the assumption of isothermal plasmas may have biased the best fit Fe abundance in groups and giant elliptical galaxies low. In this work, we take advantage of a major update of the atomic code in the spectral fitting package \textsc{spex} to re-evaluate the Fe abundance in 43 clusters, groups, and elliptical galaxies (the CHEERS sample) in a self-consistent analysis and within a common radius of 0.1$r_{500}$. For the first time, we report a remarkably similar average Fe enrichment in all these systems. Unlike previous results, this strongly suggests that metals are synthesised and transported in these haloes with the same average efficiency across two orders of magnitude in total mass. We show that the previous metallicity measurements in low temperature systems were biased low due to incomplete atomic data in the spectral fitting codes. The reasons for such a code-related Fe bias, also implying previously unconsidered biases in the emission measure and temperature structure, are discussed.
\end{abstract}





\section{Introduction}
\label{sec:intro}

The largest gravitationally bound structures in the Universe, such as giant elliptical galaxies, groups, and clusters of galaxies, are pervaded by hot, X-ray emitting atmospheres, which typically account for an important fraction (up to $\sim$50--90\%) of the total baryonic mass of these systems \citep[e.g.][]{2009ApJ...703..982G}. These hot atmospheres, hereafter defined for convenience as intra-cluster medium (ICM), are also rich in heavy elements that were produced by Type Ia and core-collapse supernovae within cluster/group members and giant central galaxies \citep[for recent reviews, see][]{2008SSRv..134..337W,2013AN....334..416D,2017AN....338..299D}. Whereas observations and simulations suggest that metals in cluster outskirts were released more than 10 Gyr ago \citep[e.g.][]{2017MNRAS.470.4583U,2017MNRAS.468..531B,2018MNRAS.476.2689B}, the epoch and origin of the enrichment in the vicinity of central galaxies is less clear.

Because the ICM is in a collisional ionisation equilibrium (CIE), abundances of various elements (typically from oxygen to nickel) can be robustly measured. This is especially true for Fe, whose both K- and L-shell transitions have high emissivities and fall within the typical energy windows ($\sim$0.5--10 keV) of our X-ray observatories.
For this reason, Fe abundances can be precisely measured in the X-ray halos of both hot, massive clusters (via the Fe-K transitions) and cooler, less massive groups and ellipticals (via the Fe-L transitions). In turn, these Fe abundance measurements are usually interpreted as a reliable tracer of the overall metallicity in clusters and groups \citep[e.g.][and references therein]{2017A&A...607A..98D}, and are thus valuable to understand the history of metal enrichment in these systems.

In the past, several works extensively studied the Fe abundance in the hot gas of either nearby ellipticals and galaxy groups \citep[e.g.][]{2005ApJ...622..187M,2006ApJ...646..143F,2011A&A...531A..15G,2014ApJ...781...36S,2014ApJ...783....8K}, or galaxy clusters \citep[e.g.][]{2004A&A...420..135T,2007A&A...465..345D,2001ApJ...551..153D,2009A&A...508..565D,2011A&A...527A.134M,2011A&A...535A..78Z}. Very few studies, however, attempted to compare directly the metal content of all these systems together \citep[e.g.][]{2010ApJ...716L..63B,2012NJPh...14d5004S}. 

In what has been perhaps the most comprehensive study so far, \citet{2017MNRAS.464.3169Y} compiled from the literature a large number of Fe abundances measured in 79 nearby groups and clusters and homogenised these measurements by extrapolating them to a radius of $r_{500}$. While in hot clusters, the Fe abundance was found to converge to a rather uniform value of $\sim0.3$ Solar, in low temperature groups and giant ellipticals the metallicity appeared to be on average significantly lower \citep[see also][]{2007MNRAS.380.1554R,2009MNRAS.399..239R}. These results were not reproduced by predictions from semi-analytical models of galaxy evolution, in which (at least) as much Fe was expected in groups as in clusters \citep{2017MNRAS.464.3169Y}.

Do theoretical models really miss some important chemodynamical process at play in galaxy groups, or do spectroscopic measurements instead suffer from unexpected biases in low-temperature systems? From an observational perspective, this question remains open. In fact, homogenising Fe abundance measurements from the literature is very challenging, essentially because: (i) different authors utilised different data reduction and analysis methods, (ii) instrumental calibration and spectral models continuously evolved with years, and (iii) the lack of accurate measurements for radial Fe profiles of individual systems out to $r_{500}$ makes the extrapolation to this radius quite uncertain. Last but not least, cooler systems ($kT \lesssim 2$ keV) require careful attention as the Fe-L complex, which is unresolved by CCD instruments, may be underestimated if one assumes the plasma to be isothermal \citep[the "Fe-bias";][]{1994ApJ...427...86B,2000MNRAS.311..176B}. Since most of the baryons (and metals) are rather in groups than in clusters, determining their accurate, unbiased metallicity is nevertheless of a crucial importance to estimate the global metal budget of the universe. Clearly, measurements of such metallicities in hot haloes at all masses need to be further investigated and better understood.

In a recent work \citep[][hereafter Paper I]{2016A&A...592A.157M}, we used \textit{XMM-Newton} EPIC observations to measure Fe -- among other elemental abundances -- in the hot haloes of 44 nearby cool-core ellipticals, groups, and clusters of galaxies (the CHEERS\footnote{CHEmical Enrichment Rgs Sample} catalog). Interestingly, we found an apparent deficit of Fe in the coolest systems, supporting the previous findings of \citet{2007MNRAS.380.1554R,2009MNRAS.399..239R} and \citet{2017MNRAS.464.3169Y}, which are in tension with theoretical expectations. In that study, however, groups and ellipticals were investigated only within 0.05$r_{500}$, making it difficult to compare with most simulations given their limited resolution. In addition, a major update of the plasma models from the \textsc{spex} fitting package \citep[][]{1996uxsa.conf..411K} has been publicly released. As briefly noted in \citet{2017A&A...603A..80M}, such an improvement could affect the Fe abundance measured by CCD instruments in cooler plasmas and potentially revise our current picture of the ICM enrichment from massive ellipticals to the largest structures of the universe.

In this Letter, we revisit the observed Fe abundances in the CHEERS sample by: (i) analysing EPIC spectra within a common astrophysical radius of 0.1$r_{500}$ -- easier to compare with simulations -- and (ii) exploring how recent spectral model improvements alter the measured Fe abundances and their interpretation. Throughout this Letter, we assume $H_0$ = 70 km s$^{-1}$ Mpc$^{-1}$, $\Omega_m$ = 0.3, and $\Omega_\Lambda$= 0.7. Error bars are given within a 68\% confidence interval. All the abundances mentioned in this work are given with respect to their proto-solar values obtained by \citet{2009LanB...4B...44L}.


\section{Reanalysis of the CHEERS sample}\label{sec:data_reduction}


The sample, data reduction, background modelling, and spectral fitting strategy are all described in detail in Paper I \citep[see also][]{2015A&A...575A..37M}. Compared to our previous work, we discard the observation of M\,89 (ObsID:0141570101) because of its high background contamination. This leaves us with \textit{XMM-Newton} EPIC observations of 43 nearby cool-core clusters, groups, and ellipticals, all being part of the CHEERS project \citep[see also][]{2015A&A...575A..38P,2017A&A...607A..98D}. The brightness of these nearby sources, combined to their relatively moderate temperature (not exceeding $\sim$8 keV), allows a robust determination of the Fe abundance with the EPIC instruments, based on the Fe-K lines and/or the Fe-L complex.

Unlike in Paper I, where the spectra were extracted within 0.05$r_{500}$ and/or 0.2$r_{500}$ (depending on the distance of the system), the goal of this paper is to measure the Fe abundance within the same physical scale. Therefore, all the spectra of our sample are re-extracted and re-analysed within 0.1$r_{500}$. The only exception is the Virgo cluster (centred on M\,87), which could be analysed only out to 0.05$r_{500}$ within the EPIC field-of-view. The redshift and hydrogen column density ($n_\text{H}$) values are adopted from Paper I.

\subsection{From \textsc{spexact} v2 to \textsc{spexact} v3}\label{sec:SPEX2vs3}

A key improvement with respect to Paper I is the updated version of the \textsc{spex} Atomic Code and Tables (hereafter \textsc{spexact}). While in Paper I our analysis relied on \textsc{spexact} v2.05 (hereafter v2), in this Letter we take advantage of the up-to-date release of \textsc{spexact} v3.04 (hereafter v3). This most recent version is the result of a major update started in 2016 (\textsc{spexact} v3.00) with further minor improvements implemented until the end of 2017 \citep{2017arXiv171205407H}. Compared to \textsc{spexact} v2, the total number of energy transitions has increased by a factor of $\sim$400, to reach more than 1.8 million in \textsc{spexact} v3. The new transitions include for instance higher principal quantum numbers for both H-like and He-like ions. In addition, significant updates were performed in collisional excitation and de-excitation rates, radiative transition probabilities, auto-ionisation and dielectronic recombination rates \citep[either from the literature or consistently calculated using the FAC\footnote{https://www-amdis.iaea.org/FAC} code][]{2008CaJPh..86..675G}. Finally, significant improvements were obtained in radiative recombination \citep{2006ApJS..167..334B,2016A&A...587A..84M} and collisional ionisation coefficients \citep{2017A&A...601A..85U}. In order to compare the effects of the improvements in a consistent way, in the following we use successively \textsc{spexact} v2 and \textsc{spexact} v3 to fit all our EPIC spectra (MOS\,1, MOS\,2, and pn are fitted simultaneously, see Paper I).

\subsection{Multi-temperature modelling}\label{sec:3T_modelling}

As already demonstrated by e.g. \citet[][Fe-bias]{1994ApJ...427...86B,2000MNRAS.311..176B} and \citet[][inverse Fe-bias]{2008ApJ...674..728R,2009A&A...493..409S}, modelling the ICM with a multi-temperature structure is essential to derive correct abundances. The most intuitive assumption would be to consider that the temperature follows a Gaussian differential emission measure distribution \cite[the \texttt{gdem} model; see e.g.][]{2006A&A...452..397D,2009A&A...493..409S}. Such a model, however, requires appreciable computing resources, especially when using \textsc{spexact} v3. A cheaper, yet still reasonable alternative would be to approximate a \texttt{gdem} distribution by modelling three temperature components (3T): (i) the main component, for which the temperature $kT_\text{mean}$ and the emission measure $Y$ are left free in the fits; (ii) a higher- and (iii) a lower-temperature components, whose temperatures $kT_\text{up}$ and $kT_\text{low}$ are left free but their $Y$ is tied to half of that of the main component. The ratio $kT_\text{up}/kT_\text{low}$ can thus be seen as the typical width of the distribution.
Although such a temperature distribution may somewhat deviate from Gaussianity in some cases, we verify that fitting (i) a subsample of systems with a \texttt{gdem} model and (ii) \texttt{gdem}-simulated EPIC spectra with a 3T model across various mean temperatures have negligible impact (always less than $\sim$6\%) on our measured Fe abundances.

Such multi-temperature modelling is particularly relevant here, as we obtain significantly better fits than when we model our spectra with a single-temperature component only. Moreover, in addition to the fact that all our systems are classified as cool-core, they are also known to exhibit clear temperature gradients within $\sim$0.1$r_{500}$ \citep[see e.g. results from the ACCEPT catalog][]{2009ApJS..182...12C}.


\section{Results}\label{sec:results}

\begin{figure}
        \centering
                \includegraphics[trim=0cm 0cm 0.5cm 0.4cm, clip=true,width=\columnwidth]{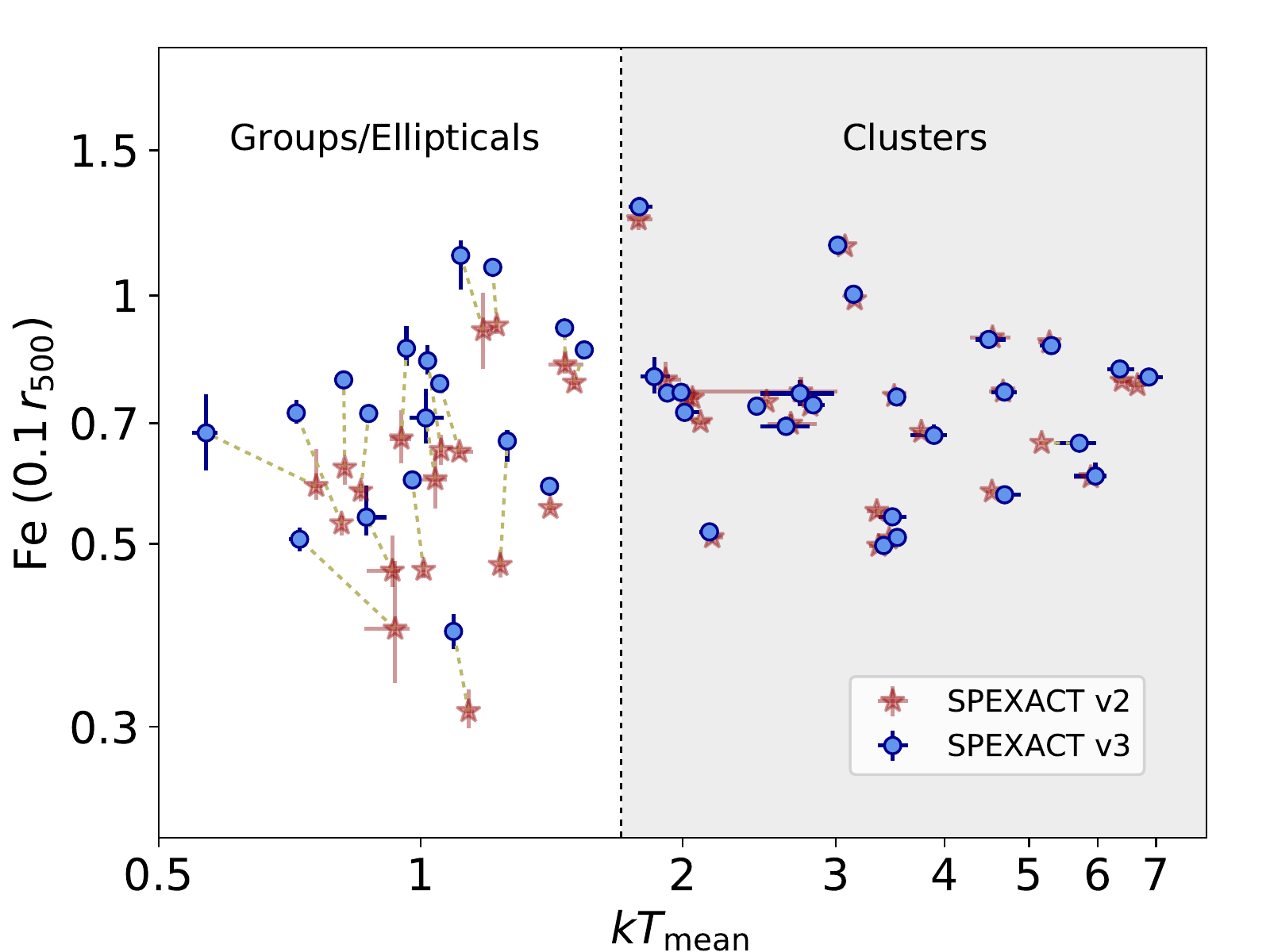} 

        \caption{Iron abundance measured as a function of the mean temperature within 0.1$r_{500}$ of the ellipticals, galaxy groups and clusters from the CHEERS sample (only M\,89 is discarded, see text). For a given system, the corresponding \textsc{spexact} v2 (orange stars) and \textsc{spexact} v3 (blue dots) measurements, both obtained using a 3T model (see text), are tied by a green-brown dashed line. Clusters and groups/ellipticals are delimited arbitrarily beyond and below $kT_\text{mean} = 1.7$ keV, respectively.}
\label{fig:kT_Fe}
\end{figure}

The measured Fe abundances of the 43 CHEERS systems reanalysed within 0.1$r_{500}$ are shown as a function of their $kT_\text{mean}$ in Fig.~\ref{fig:kT_Fe}. Because the overall temperature of the ICM scales with the total mass $M$ of the system as $\sim$$M^{2/3}$ \citep[e.g.][]{2013SSRv..177..247G}, $kT_\text{mean}$ can be seen as a reasonable proxy for the total mass of our sources. Therefore, we split our sample into two subsamples, namely: (i) "clusters", for which $kT_\text{mean} > 1.7$ keV, and (ii) "groups/ellipticals", for which $kT_\text{mean} < 1.7$ keV. The choice of the threshold value $kT_\text{mean} = 1.7$ keV is of course arbitrary, but well justified by the usual classification attributed to each system in the literature.

\begin{figure}
        \centering
                \includegraphics[trim=0cm 0cm 0.5cm 0.4cm, clip=true,width=\columnwidth]{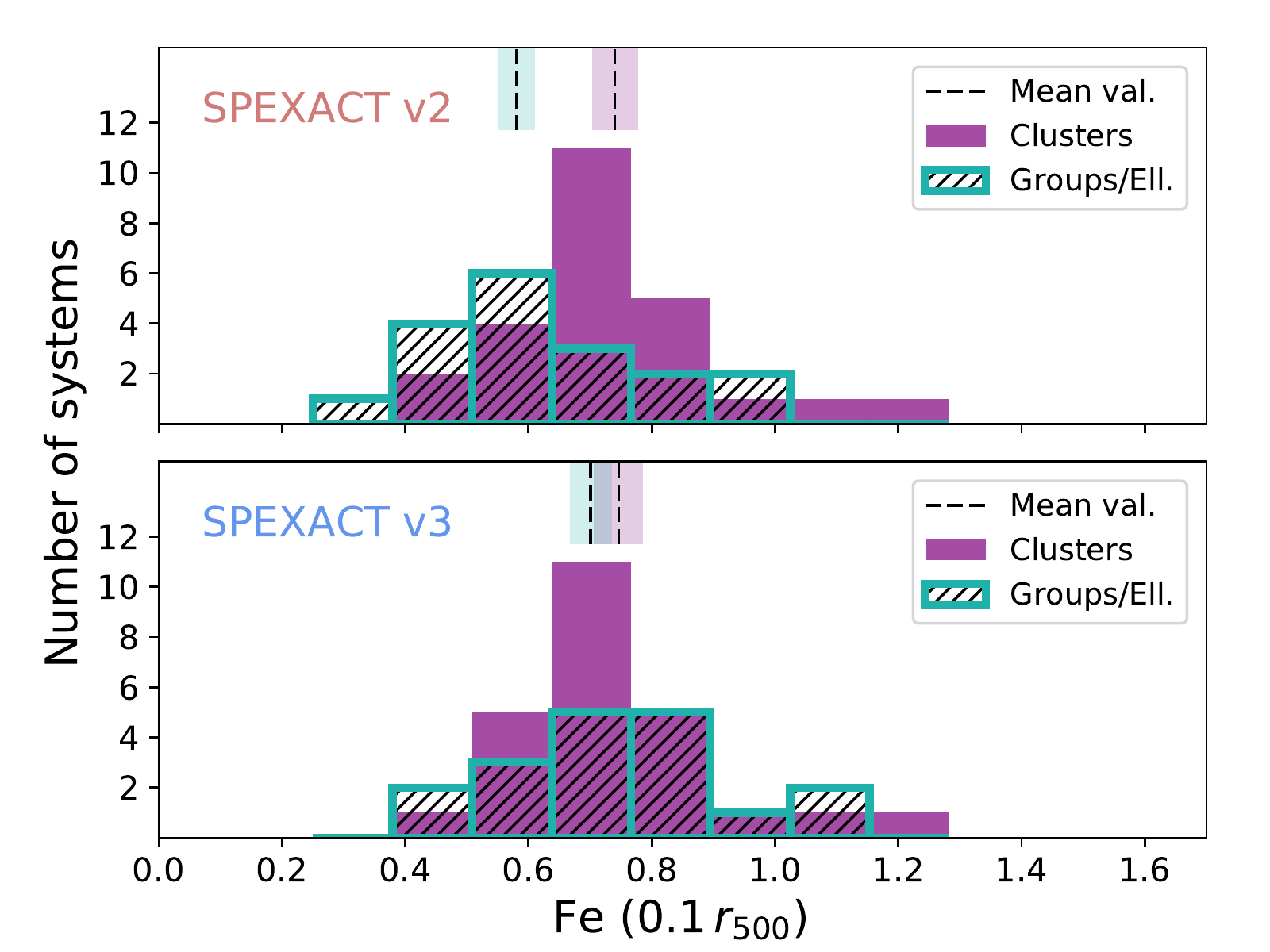} 

        \caption{Histograms showing the Fe abundance distribution of the CHEERS sample, when using successively \textsc{spexact} v2 (\textit{top panel}) and \textsc{spexact} v3 (\textit{bottom panel}). In each case, the distribution for clusters ($kT_\text{mean} > 1.7$ keV) and groups/ellipticals ($kT_\text{mean} < 1.7$ keV) is shown separately. The mean value of each distribution (and corresponding errors) is shown by the vertical dashed lines (and filled areas around them).}
\label{fig:kT_Fe_hist}
\end{figure}

While compared to \textsc{spexact} v2,  the Fe abundances measured in clusters remain essentially unchanged, the Fe abundances in groups and ellipticals are systematically revised upwards when using \textsc{spexact} v3. This result is better quantified in Fig.~\ref{fig:kT_Fe_hist}, where the distribution of Fe abundances is compared between clusters and groups/ellipticals, using the two versions of the code. Based on the entire sample, the \textsc{spexact} v3 results provide a mean Fe abundance of $0.74 \pm 0.03$ with an intrinsic scatter of 25\% (computed following the method described in Paper I). When splitting the sample, we find consistent average Fe abundances of $0.75 \pm 0.04$ and $0.70 \pm 0.03$ for clusters and groups/ellipticals, respectively. This is in contrast with the \textsc{spexact} v2 results, where the average Fe abundance values for clusters ($0.75 \pm 0.04$) and groups/ellipticals ($0.58 \pm 0.03$) are significantly different.
In other words, spectral fits obtained using updated atomic data indicate that the average concentration of Fe in the hot haloes of groups and giant ellipticals is the same as that in clusters of galaxies.

Systems for which $kT_\text{mean}$ lies within 2--3 keV exhibit both Fe-L and Fe-K lines, hence their Fe abundance can be constrained by each of these two features separately. We test this approach on M\,87 and EXO\,0422, both having good data quality. Compared to the Fe estimated from the "full band" fits, we find that the <2 keV local fits (Fe-L lines only) provide negligible biases (-2\% and +6\% for M\,87 and EXO\,0422, respectively). These biases become somewhat larger (respectively +15\% and -12\%) in the >2 keV local fits (Fe-K lines only); however they are not systematic and remain well below the typical 25\% scatter reported above. Although this issue is well known \citep{2008ApJ...674..728R,2009A&A...493..409S} and concerns less than $\sim$14\% of our systems, this mismatch will deserve attention with future high-resolution spectroscopy missions.

\subsection{The code-related Fe-bias}\label{sec:Fe_bias}

\begin{figure}
        \centering
                \includegraphics[trim=0cm 0.2cm 0.5cm 0.4cm, clip=true,width=\columnwidth]{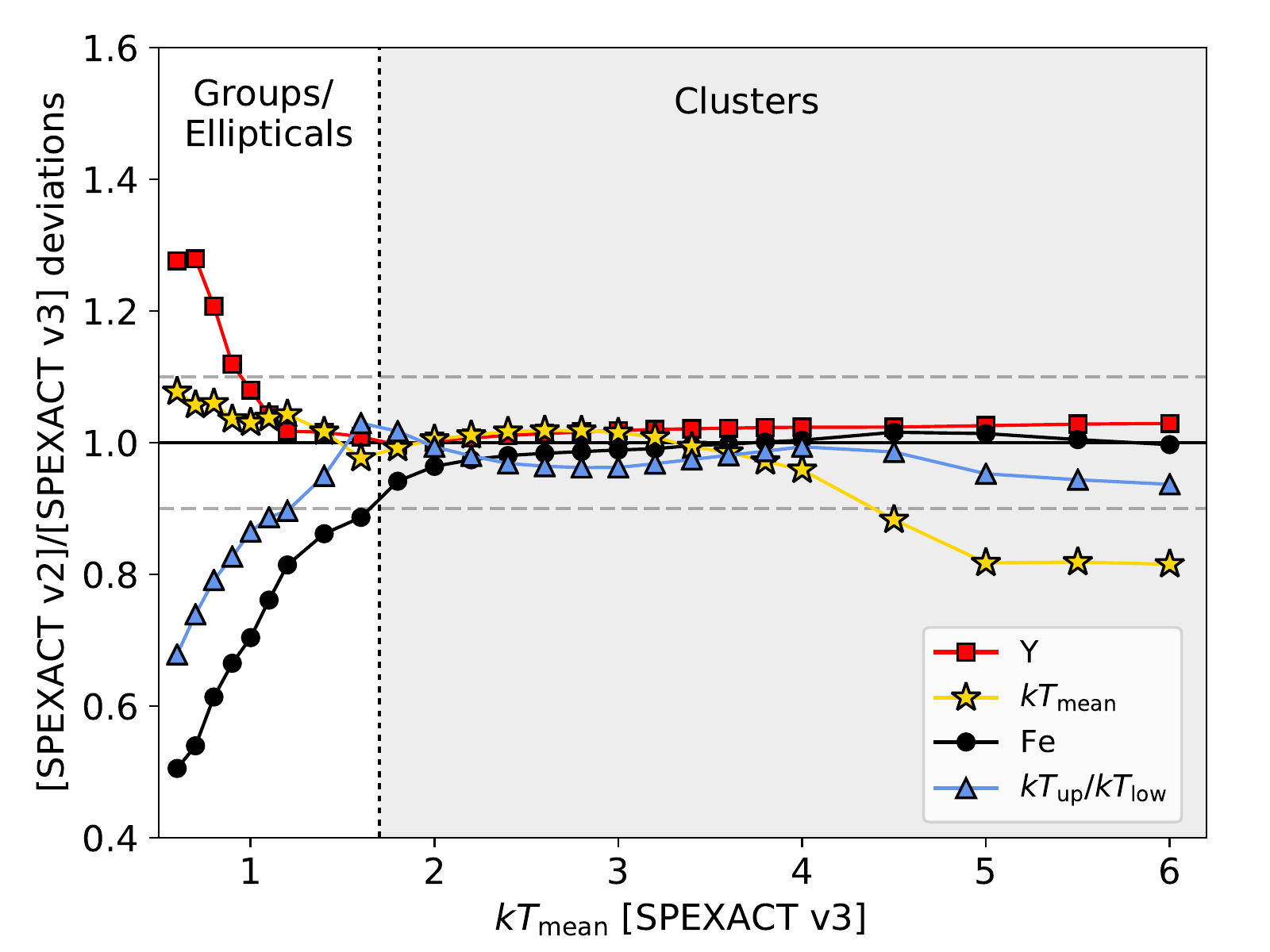} 

        \caption{Relative deviations on the parameters $Y$, $kT_\text{mean}$, Fe, and the ratio $kT_\text{up}/kT_\text{low}$ when EPIC mock spectra of 3T plasma (simulated using \textsc{spexact} v3 for various initial mean temperatures) are fitted using \textsc{spexact} v2. The two horizontal dashed lines indicate the $\pm$10\% relative deviations.}
\label{fig:SPEX2vs3_3T}
\end{figure}

In clusters, the Fe abundance determination is predominantly based on the prominent Fe-K lines. Since only the low temperature groups/ellipticals are significantly affected by the update of \textsc{spexact}, the reason for such a change is to be found in the Fe-L emission, which is dominant at $kT_\text{mean} \lesssim 2$ keV. In order to better understand the code-related Fe bias that we report above, we adapt an instructive exercise previously introduced in \citet{2017A&A...603A..80M} and \citet{2017A&A...607A..98D}. In short, we start by using \textsc{spexact} v3 to simulate mock EPIC spectra with 100 ks exposure on a grid of various $kT_\text{mean}$ values. In all these simulations $Y$ and the abundances are assumed to be $10^{72}$ m$^{-3}$ and 1 proto-solar, respectively. Moreover, $kT_\text{up}$ and $kT_\text{low}$ are assumed such that $kT_\text{up}/kT_\text{low} = 2.8$. As a second step, we fit these mock spectra using \textsc{spexact} v2 with $Y$, Fe, $kT_\text{mean}$, $kT_\text{up}$, and $kT_\text{low}$ as free parameters. The relative deviation of these \textsc{spexact} v2 best-fit parameters with respect to their input \textsc{spexact} v3 values is shown in Fig.~\ref{fig:SPEX2vs3_3T} as a function of the input mean temperature. As expected from our results above, the Fe consistency between the two versions of \textsc{spexact} is excellent in the clusters regime, while it dramatically deteriorates when the plasma becomes cooler than $\sim$2 keV. In addition, other interesting effects occur in the groups/ellipticals regime. Below $kT_\text{mean} \lesssim 1.5$ keV and $kT_\text{mean} \lesssim 1$ keV, the ratio $kT_\text{up}/kT_\text{low}$ and $Y$ are respectively under- and overestimated by \textsc{spexact} v2. The mean temperature, however, remains reasonably reproduced by \textsc{spexact} v2, except for very hot plasmas where $kT_\text{mean}$ is at most $\sim$15\% underestimated (though without affecting the Fe abundance).

\begin{figure}
        \centering
                \includegraphics[trim=0.2cm 0.7cm 0.1cm 0.6cm, clip=true,width=\columnwidth]{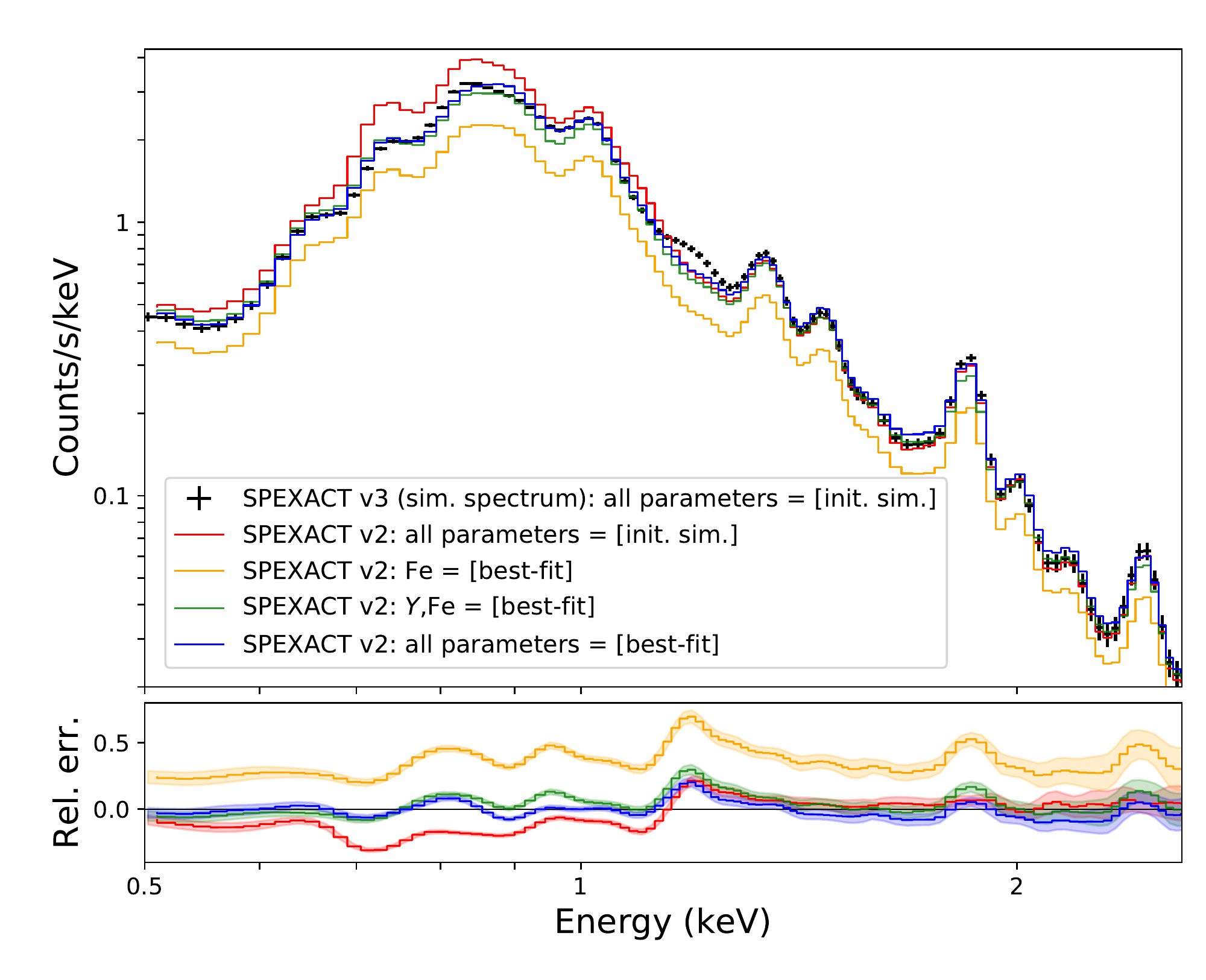}

        \caption{EPIC MOS\,2 simulated spectrum of a 3T plasma with $kT_\text{mean} = 0.7$ keV, using \textsc{spexact} v3. For comparison, we show the same model calculated using \textsc{spexact} v2 (red). Then, we progressively fix the Fe (orange), $Y$ (green), and eventually $kT_\text{mean}$, $kT_\text{up}$, and $kT_\text{low}$ (blue) to their \textit{a posteriori} best-fit \textsc{spexact} v2 values. The residuals of such models with respect to the input simulated spectrum are shown in the bottom panel.}
\label{fig:Fe-L_complex}
\end{figure}

To better understand all the biases we observe in cool plasmas with the EPIC instruments, we investigate further the case of a 3T plasma simulated for 100 ks with \textsc{spexact} v3, assuming $kT_\text{mean} = 0.7$ keV (Fig.~\ref{fig:Fe-L_complex}, black data points). A direct comparison of this simulated spectrum with its equivalent model using \textsc{spexact} v2 (Fig.~\ref{fig:Fe-L_complex}, red line) shows significant discrepancies throughout the entire Fe-L complex (0.6--1.2 keV). In fact, the emissivity of many important lines (e.g. Fe XVII at $\sim$0.73 keV; Fe XVIII at $\sim$0.77 keV) were revised lower with the update of \textsc{spexact}, while new transitions were incorporated and/or updated with a higher emissivity (e.g. Fe XVIII at $\sim$1.18 keV). When fixing the Fe abundance to its best-fit value estimated \textit{a posteriori} by \textsc{spexact} v2 (Fig.~\ref{fig:Fe-L_complex}, orange line), the emitting bump at $\sim$0.7 get smoother, in better agreement with the overall shape of the Fe-L complex. However, over the entire soft band the flux significantly decreases, which the fit attempts to "correct" by increasing $Y$ (Fig.~\ref{fig:Fe-L_complex}, green line). Finally, the fit smooths the residual bumps (in particular around $\sim$0.9--1 keV) by simultaneously decreasing $kT_\text{up}$ and increasing $kT_\text{low}$ to provide a formally acceptable -- but incorrect -- best-fit to the input spectrum (Fig.~\ref{fig:Fe-L_complex}, blue line).

In summary, in cool plasmas the emission measure, the Fe abundance, and the width of the temperature distribution ($kT_\text{up}/kT_\text{low}$) influence each other to reproduce the observed shape of the unresolved Fe-L complex. As a consequence, even outdated spectral codes can reasonably fit the Fe-L complex, yet providing strongly biased measurements. This conspiracy between all these parameters explain the code-related Fe-bias that we report in this Letter.

\section{Implications for the iron content in ellipticals, groups and clusters}\label{sec:discussion}


By measuring Fe abundances within 0.1$r_{500}$ in a self-consistent way and using the latest \textsc{spexact} version available to date, we report for the first time similar Fe abundances in ellipticals, galaxy groups, and galaxy clusters. In other words, gas-phase metallicities remain constant across two orders of magnitude in halo mass.

These new results contradict previous papers \citep[e.g.][]{2009MNRAS.399..239R,2010ApJ...716L..63B,2012NJPh...14d5004S,2017MNRAS.464.3169Y}, which reported systematically lower Fe abundances in groups and/or ellipticals with respect to the hotter clusters of galaxies (although \citet{2014ApJ...783....8K} reported similar average Fe abundances as reported here, albeit for ellipticals only). Rather than \textsc{spex}, most of those previous studies used many (very different) versions of \textsc{apec} to fit their data, making a direct comparison with this work difficult. All these (mostly outdated) atomic codes, however, likely encountered similar problems of a too simplistic modelling of the Fe-L transitions. From a theoretical perspective, that trend was not trivial to explain. For example, when comparing the observational trend with a semi-analytic model, \citet{2017MNRAS.464.3169Y} did not succeed to reproduce the previously reported positive temperature-metallicity correlation in galaxy groups. Instead, the metal content in low-mass systems is systematically overestimated by their model.

Our present results have interesting consequences, in particular given that the investigated systems exhibit very different stellar- to ICM-mass fractions. Because this fraction is lower in rich clusters than in less massive systems, invariant Fe abundances could be explained only if the effective ICM enrichment considerably increases with the mass of the system. Such requirements have been difficult to reconcile with the observed stellar populations in clusters so far \citep[e.g.][]{2013ApJ...773...52L,2014MNRAS.444.3581R}. The story, however, is different if the Fe present in the ICM is unrelated to the current stellar population of these systems. In addition to the increasing evidence towards an early ICM enrichment in cluster outskirts \citep[e.g.][]{2013Natur.502..656W,2017MNRAS.469.1476S,2017MNRAS.470.4583U}, central Fe peaks were also found to be in place already at $z \sim 1$ \citep{2014A&A...567A.102D,2017MNRAS.472.2877M} and exhibit the same radial distribution as SNcc products \citep{2017A&A...603A..80M}. These recent findings suggest that recent SNIa explosions and stellar mass loss from central galaxies do not significantly contribute to the ICM enrichment.

In this context, the similar Fe abundances found in hot haloes spanning different mass ranges constitute an additional support toward this early enrichment scenario, even in their central parts. Since they grow hierarchically, isolated massive ellipticals and assembling groups can be seen as the first steps of the formation of more massive clusters. Although, admittedly, nearby groups may have different specific properties (star formation, AGN feedback, etc.) than high-redshift proto-clusters, the mass-invariance of Fe abundances at low redshift suggests that the bulk of metals in hot haloes was already in place well before clusters effectively assembled. Our new measurements are directly confronted to (and are found to be in good agreement with) recent chemo- and hydrodynamical simulations in a companion paper \citep{2018arXiv180306297T}, to which we refer the reader for a more detailed discussion.

We remind that these integrated measurements cover 0.1$r_{500}$, without further information on their inner or outer spatial distributions. The question of whether clusters and groups/ellipticals are really self-similar in terms of metal enrichment would require at least to derive the individual abundance profiles for the entire sample using \textsc{spexact} v3 \citep[for a similar work using \textsc{spexact} v2, see][]{2017A&A...603A..80M}. Because of the non-negligible time required by \textsc{spexact} v3 to fit each spectrum, we leave such a study for future work.

In addition to the code-related Fe bias discussed in this work, we also note from Fig.~\ref{fig:SPEX2vs3_3T} that fitting the spectra of cool systems with an outdated plasma code may also bias the emission measure, the mean temperature and the $kT_\text{up}/kT_\text{low}$ ratio by +35\%, +7\%, and -24\%, respectively. In turn, these biases may have consequences on the estimates of further interesting quantities. For instance, we estimate that the ICM pressure, usually defined as $P = n_e kT$, can be biased high by $\sim$19\% in the case of a $\sim$0.7 keV plasma. Unlike the pressure, the ICM entropy, usually defined as $K = kT/n_e^{2/3}$, remains very close to its true value, with a underestimate of less than $\sim$1\%. Similarly, the total hydrostatic mass is not expected to be affected by more than a few percent, as temperature and density gradients do not change dramatically. A more precise quantification, however, is left to future work. Our results also reveal the complication of measuring accurately the temperature structure of lower-mass systems, as long as the Fe-L complex remains unresolved by the observing instruments.

Finally, it should be reminded that no spectral code is perfect. It is certain that further improvements on \textsc{spexact} will be pursued in the future, with potential implications on the interpretation of moderate resolution spectra of X-ray sources. In that respect, micro-calorimeters onboard future missions such as \textit{XARM} and \textit{Athena} will enable us to observe the Fe-L complex with unprecedented resolution. These observations will be invaluable to better understand all the radiation processes in the ICM and push our knowledge of astrophysical plasma emission to the next level.

\section*{Acknowledgements}

The authors thank the referee for their constructive comments that helped to improve the manuscript as well as Kiran Lakhchaura for fruitful discussions. F.M. is supported by the Lend\"ulet LP2016-11 grant awarded by the Hungarian Academy of Sciences. This work is partly based on the \textit{XMM-Newton} AO-12 proposal ``\emph{The XMM-Newton view of chemical enrichment in bright galaxy clusters and groups}'' (PI: de Plaa), and is a follow-up of the CHEERS (CHEmical Evolution Rgs cluster Sample) collaboration; the authors thank all its members. This work is based on observations obtained with \textit{XMM-Newton}, an ESA science mission with instruments and contributions directly funded by ESA member states and the USA (NASA). The SRON Netherlands Institute for Space Research is supported financially by NWO, the Netherlands Organisation for Scientific Research.

\bibliography{Letter_I_Fe_final}{}
\bibliographystyle{mnras}




\end{document}